\documentclass[prb,aps,twocolumn,showpacs]{revtex4-2} 
\usepackage{amsmath,amssymb,amsthm}
\usepackage{graphicx}
\usepackage{subfigure}
\usepackage{amsmath}
\usepackage{amsfonts,amssymb}
\usepackage{bm}
\usepackage{float}
\usepackage{color}
\usepackage{sidecap}
\allowdisplaybreaks
\usepackage{soul}
\setstcolor{red}
\usepackage[normalem]{ulem}
\usepackage{hyperref}
\hypersetup{colorlinks=true,breaklinks,urlcolor=blue,linkcolor=blue,citecolor=blue}

\begin{document}

\title{Coexistence of non-Hermitian skin effect and extended states in  one-dimensional nonreciprocal lattices}
\author{Han Xiao}
\affiliation{Department of Physics, Capital Normal University, Beijing 100048, China}
\author{Qi-Bo Zeng}
\email{zengqibo@cnu.edu.cn}
\affiliation{Department of Physics, Capital Normal University, Beijing 100048, China}

\begin{abstract}
 We study the one-dimensional non-Hermitian lattices with staggered onsite modulations and nonreciprocal hopping up to the next-nearest-neighboring (NNN) sites. Due to the NNN nonreciprocity, the non-Hermitian skin effect (NHSE) in the system under open boundary conditions (OBC) can be energy dependent, and there will be NHSE edges in the eigenenergy spectrum, which separates the eigenstates localized at the opposite ends of the lattice. We find that the interplay between the nonreciprocal hopping and onsite modulations can reverse the direction of the skin effect and modify the position of the NHSE edge. Moreover, by tuning the system parameters, some of the eigenstates under OBC will become fully extended with the corresponding eigenenergies being imaginary under both open and periodic boundary conditions. Thus, the extended states can coexist with the NHSH in the same system. The NHSE can even be completely dissolved with all the eigenstates being extended when the modulation is imaginary. Our work unveils the intricate interplay between onsite modulations and nonreciprocal hopping in non-Hermitian systems.    
\end{abstract}
\maketitle
\date{today}

\section{Introduction}
It is well known that in conventional quantum mechanics, the model Hamiltonian of the system under study is required to be Hermitian such that the reality of eigenenergies can be guaranteed. However, in the 1990s, Bender et al. found that Hamiltonians that are not Hermitian but $\mathcal{PT}$ symmetric can also have real eigenenergies~\cite{Bender1998PRL,Bender2002PRL,Bender2007RPP}. The existence of real spectrum is further extended to pseudo-Hermitian Hamiltonians~\cite{Mostafazadeh2002JMP,Mostafazadeh2010IJMMP,Moiseyev2011Book,Zeng2020PRB1,Kawabata2020PRR,Zeng2021NJP}. During the past few decades, non-Hermitian physics has undergone rapid developments~\cite{Cao2015RMP,Konotop2016RMP,Ganainy2018NatPhy,Ashida2020AiP,Bergholtz2021RMP}. A plethora of non-Hermitian Hamiltonians have been widely exploited to effectively describe both classical~\cite{Makris2008PRL,Klaiman2008PRL,Guo2009PRL,Ruter2010NatPhys,Lin2011PRL,Regensburger2012Nat,Feng2013NatMat,Peng2014NatPhys,Wiersig2014PRL,Hodaei2017Nat,Chen2017Nat} and quantum open systems~\cite{Brody2012PRL,Lee2014PRX,Li2019NatCom,Kawabata2017PRL,Hamazaki2019PRL,Xiao2019PRL,Wu2019Science,Yamamoto2019PRL,Yamamoto2019PRL,Naghiloo2019NatPhys,Matsumoto2020PRL,Mandal2020PRL, Mandal2022ACS}.

Apart from the exotic spectral properties, such as the exceptional points in non-Hermitian systems, another unique phenomenon is the non-Hermitian skin effect (NHSE), where a macroscopic number of eigenstates accumulate at the system boundaries~\cite{Yao2018PRL1,Yao2018PRL2}. The NHSE can influence the system significantly and there has been a recent burst of studies on it~\cite{Alvarez2018PRB,Alvarez2018EPJ,Lee2019PRB,Zhou2019PRB,Kawabata2019PRX,Song2019PRL,Okuma2020PRB,Xiao2020NatPhys,Yoshida2020PRR,Longhi2019PRR,Yi2020PRL,Claes2021PRB,Haga2021PRL,Zeng2022PRA,Zeng2022PRB}. For example, the band topology and the conventional bulk-boundary correspondence principle in topological phases might break down in systems with NHSE~\cite{Yao2018PRL1,Yao2018PRL2,Kunst2018PRL,Jin2019PRB,Yokomizo2019PRL,Herviou2019PRA,Zeng2020PRB,Borgnia2020PRL,Yang2020PRL2,Zirnstein2021PRL,Zhang2022arxiv}. Moreover, the spectra of systems with NHSE can be very sensitive to the change of boundary conditions~\cite{Xiong2018JPC}, which are applicable in designing new quantum sensors~\cite{Budich2020PRL,Koch2022PRR}. The presence of NHSE in the system under open boundary conditions (OBCs) is closely related to the point gap in the complex energy spectrum under periodic boundary conditions (PBCs)~\cite{Okuma2020PRL,Zhang2020PRL}.

Generally, in systems with nonreciprocal or asymmetric hopping, such as the paradigmatic Hatano-Nelson model with constant nonreciprocity~\cite{Hatano1996PRL}, all the eigenstates are moved to the boundaries due to the NHSE. However, the states can still stay inside the bulk under certain circumstances. For example, in systems with random or quasiperiodic disorders, the eigenstates can be localized inside the bulk due to the Anderson localization phenomenon, where the NHSE cannot move these states to the boundaries when the disorders are strong enough~\cite{Shnerb1998PRL,Gong2018PRX,Jiang2019PRB,Zeng2020PRR,Liu2021PRB1,Liu2021PRB2}. Recently, it has also been found that in systems with linearly increasing nearest-neighboring nonreciprocal hopping, the eigenstates can be immune to the NHSE and become tightly bound states inside the bulk as the nonreciprocity increases~\cite{Zeng2024PRB}. It will be interesting to ask whether it is possible to get rid of the NHSE and obtain extended states inside the non-Hermitian lattices with nonreciprocal hopping. 

To answer the above question, we study the one-dimensional (1D) lattices with nonreciprocal hopping and staggered onsite modulations. The nonreciprocal hoppings are introduced both in the nearest-neighboring (NN) and NNN hopping terms. We find that the NHSE under OBC can be energy dependent, where NHSE edges emerge in the spectrum and separate the eigenstates localized at the opposite ends of the lattice. The interplay between the on-site modulations and the nonreciprocal hopping leads to significant changes in the spectral properties and can reverse the direction of the NHSE. Most interestingly, some of the eigenstates will become extended by tuning the system parameters, with the corresponding eigenenergies being purely imaginary under both OBC and PBC. Other eigenstates, however, are still localized at the ends of the lattice due to the NHSE. Thus, the extended states can coexist with the NHSE in such systems. If the modulations are imaginary, which represent the onsite physical gain and loss, the NHSE can even be dissolved completely. And all the eigenstates will be extended when the modulation is stronger than a critical value. Our work unveils the exotic properties of the non-Hermitian systems with NNN nonreciprocal hopping and its intricate interplay with on-site modulations.

The rest of the paper is organized as follows. In Sec.~\ref{Sec2}, we will first introduce the model Hamiltonian of the 1D lattices with nonreciprocal hopping and onsite modulations. In Sec.~\ref{Sec3}, we discuss the variations of the eigenenergy spectrum and the NHSE under the influences of onsite modulations and nonreciprocal hopping. Then we will further investigate the emergence of extended states and its coexistence with NHSE in such systems in Sec.~\ref{Sec4}. The influences of imaginary on-site modulations are discussed in Sec.~\ref{Sec5}. Finally in Sec.~\ref{Sec6}, we will summarize our results. 

\begin{figure}[t]
	\includegraphics[width=3.4in]{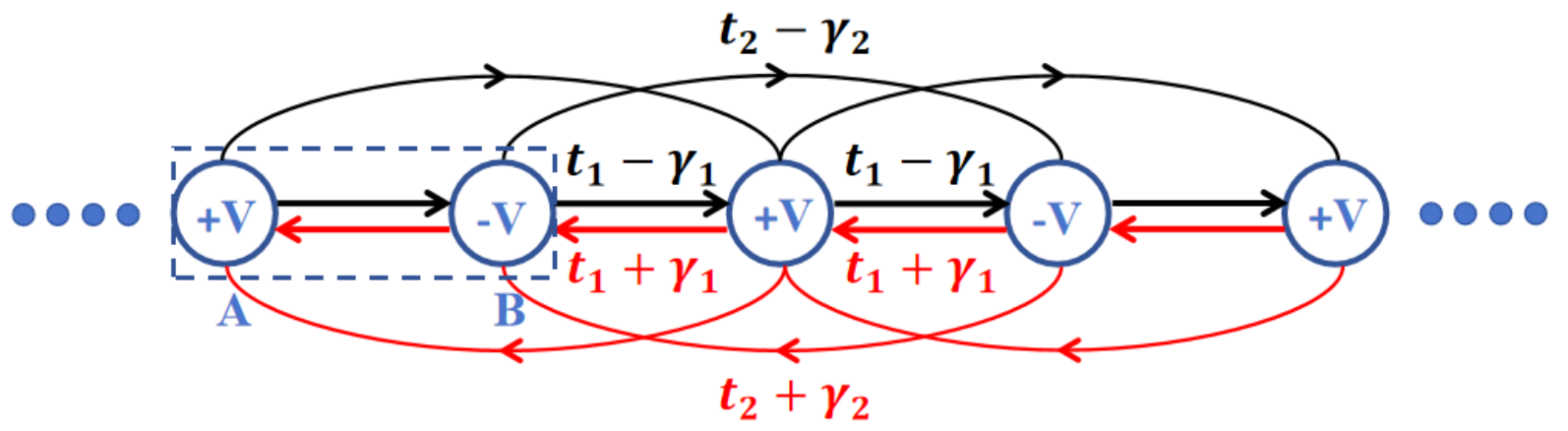}
	\caption{(Color online) Schematic of the 1D lattice with staggered on-site modulations and nonreciprocal hoppings up to the next-nearest-neighboring sites, which are represented by $\pm V$, $(t_1\pm \gamma_1)$, and $(t_2\pm \gamma_2)$, respectively. The dashed square indicates the unit cell, which is composed of sites $A$ and $B$. The onsite modulation $V$ can be real or imaginary.}
	\label{fig1}
\end{figure}

\section{Model Hamiltonian}\label{Sec2}
We introduce the 1D lattice with staggered onsite modulations and nonreciprocal hoppings up to the next-nearest-neighboring (NNN) sites. Figure~\ref{fig1} shows the schematic illustration of the lattice under OBC, which is described by the following model Hamiltonian
\begin{widetext}
\begin{equation}\label{H}
	\begin{aligned}
	H &= \sum_{j=1}^{N} \left(V c_{jA}^\dagger c_{jA} - V c_{jB}^\dagger c_{jB} \right)\\ 
	&+ \sum_{j=1}^{N-1} \left[ (t_1 + \gamma_1) ( c_{jA}^\dagger c_{jB} + c_{jB}^\dagger c_{j+1,A} ) + (t_1 - \gamma_1) ( c_{jB}^\dagger c_{jA} + c_{j+1,A}^\dagger c_{jB} ) \right] \\
	&+ \sum_{j=1}^{N-2} \left[ (t_2 + \gamma_2) ( c_{jA}^\dagger c_{j+1,A} + c_{jB}^\dagger c_{j+1,B} ) + (t_2 - \gamma_2) ( c_{j+1,A}^\dagger c_{jA} + c_{j+1,B}^\dagger c_{jB} ) \right],
	\end{aligned}
\end{equation}
where $c_{jA}$ and $c_{jB}$  ($c_{jA}^\dagger$ and $c_{jB}^\dagger$) are the annihilation (creation) operators of spinless fermions at the $A$ and $B$ site in the $j$th unit cell. The forward and backward hopping between the nearest-neighboring (NN) and NNN sites are represented by ($t_s-\gamma_s$) and ($t_s+\gamma_s$) with $(s=1,2)$, respectively. $t_s$ and $\gamma_s$ are real numbers. $V$ is the onsite potential which can be real or imaginary. If $V=0$, then the model reduces to the one studied in Ref.~\cite{Zeng2022PRB2}, where the NHSE is energy dependent and there is an NHSE edge in the energy spectrum separating the states localized at opposite ends of the lattice. $N$ denotes the number of unit cells and the lattice size is $L=2N$. Throughout this paper, we will take $t_1=1$ as the energy unit. The eigenenergies and eigenstates under OBC can be obtained by diagonalizing the above Hamiltonian numerically.

By transforming the Hamiltonian into the momentum space, we get
\begin{equation}\label{Hk}
	H(k) = \sum_k \begin{bmatrix} c_{kA}^\dagger & c_{kB}^\dagger \end{bmatrix}
	\begin{bmatrix}
		2 ( t_2 \cos k + i \gamma_2 \sin k ) + V & (t_1 - \gamma_1) e^{-ik} + (t_1 + \gamma_1) \\
		(t_1 + \gamma_1) e^{ik} + (t_1 - \gamma_1) & 2 ( t_2 \cos k + i \gamma_2 \sin k ) - V
	\end{bmatrix}
	\begin{bmatrix} c_{kA} \\ c_{kB} \end{bmatrix}.
\end{equation}
Then we can obtain the PBC energy spectrum as 
\begin{equation}\label{Ek}
	\begin{aligned}
	E_{\pm}(k) &= 2 \left( t_2 \cos k + i \gamma_2 \sin k \right) \pm \sqrt{4 \left(t_1 \cos \frac{k}{2} + i \gamma_1 \sin \frac{k}{2} \right)^2 + V^2} \\
	&= 2 \left( t_2 \cos k + i \gamma_2 \sin k \right) \pm \sqrt{2(t_1^2 - \gamma_1^2) + 2(t_1^2 + \gamma_1^2) \cos k + 4 i t_1 \gamma_1 \sin k + V^2}.
	\end{aligned}
\end{equation}

Under OBC, the eigenstates will be shifted to the ends of the lattice due to the NHSE. In order to distinguish the states localized at the opposite ends, we introduce the directional IPR (dIPR) as~\cite{Zeng2022PRB}
\begin{equation}
	\text{dIPR} (\Psi_n) = \mathcal{P}(\Psi_n) \sum_{j=1}^L \frac{|\Psi_{n,j}|^4}{(\langle \Psi_n | \Psi_n \rangle)^2}, 
\end{equation}
with $\mathcal{P}(\Psi_n)=sgn \left[ \sum_{j=1}^L \left(  j- \frac{L}{2} - \delta \right) |\Psi_{n,j}| \right]$. Here $\delta \in (0,0.5)$ is a constant. $sgn(x)$ takes the sign of the argument, which is positive (negative) for $x>0$ ($x<0$). The dIPR is positive when $\Psi_n$ is localized at the right end but is negative when $\Psi_n$ is localized at the left end. If $\text{dIPR} \rightarrow 0$, then the state is extended.

\end{widetext}

In the following, we will mainly focus on the systems with real onsite modulations, i.e., when $V$ is a real number. If $V$ is imaginary, which represents the onsite gain and loss in the lattice, we will show that similar phenomena will also happen.  

\begin{figure}[t]
	\includegraphics[width=3.4in]{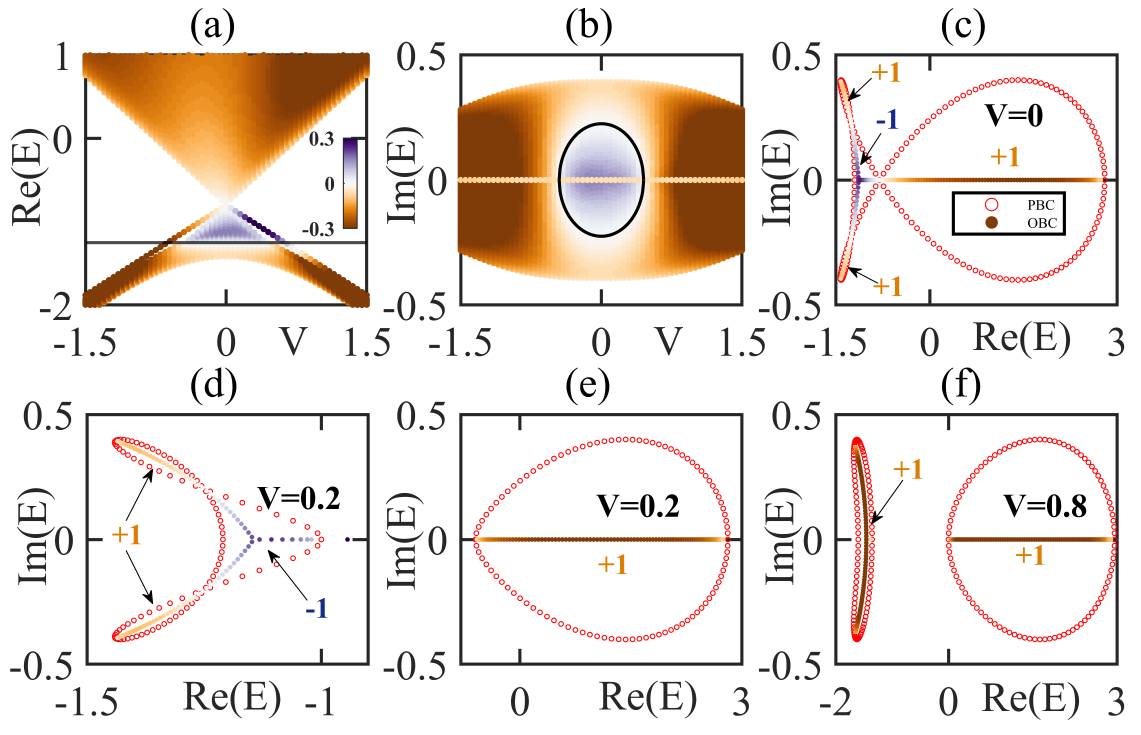}
	\caption{(Color online) Eigenenergy spectrum for the lattices with $\gamma_1=0$. Panels (a) and (b) show the real and imaginary parts of the OBC spectrum as a function of the on-site modulation $V$. The black line and circle are the NHSE edge that separates the states localized at the opposite ends. Panels (c)-(f) are the spectra under different values of $V$: (c) $V=0$, (d), (e) $V=0.2$ (the two bands are plotted independently for clarity), and (f) $V=0.8$. The colorbar indicates the dIPR values of the eigenstates under OBC and the red dots represent the PBC spectra. The numbers $\pm 1$ represent the winding number of the loops formed by the PBC eigenenergies. Other parameters: $t_1=1$, $t_2=0.4$, $\gamma_2=0.2$, and $N=100$.}
	\label{fig2}
\end{figure}

\section{Direction reversal of the NHSE}\label{Sec3}
Due to the presence of nonreciprocal hopping between the NNN sites, the NHSE will become energy-dependent, and NHSE edge emerges in the energy spectrum, separating the eigenstates localized at opposite ends of the 1D lattice~\cite{Zeng2022PRB2}. In this section, we will check how the onsite modulations will modify the behaviors of the NHSE. We start with a simple case by setting $\gamma_1=0$, implying that the nonreciprocity only exists in the NNN hopping terms. Figures~\ref{fig2} (a) and \ref{fig2}(b) show the real and imaginary parts of the eigenenergy spectrum under OBC as a function of $V$. The colorbar indicates the dIPR values of the eigenstates. For the eigenstates localized at the left (or right) end, we have $\text{dIPR} (\Psi_n)<0$ $ \left[\text{or dIPR}(\Psi_n)>0 \right] $, which are represented by the brown (or blue) dots in the figures. As we can see from the OBC spectrum, when there is no on-site modulation, i.e., $V=0$, there is only one band. When $V$ becomes nonzero, the real part will be split into two bands by a gap, in correspondence with the staggered onsite modulations. We label the band with larger (or smaller) real parts by $E_{+}$ (or $E_{-}$). From the dIPR values, we find that the eigenstates in the band $E_{+}$ are all localized at the left end while, for those in the band $E_{-}$, some of the eigenstates are shifted to the right end when $V$ is small. Moreover, for the system with $V=0$, NHSE edge already shows up. As $V$ increases, the blue region in Fig.~\ref{fig2}(a) shrinks, implying that the number of eigenstates localized at the right end will decrease. The NHSE edge can be determined by calculating the self-intersection point in the PBC spectrum (red dots). As $V$ increases further, all the eigenstates localize at the left end and there will be no NHSE edge in the spectrum.

\begin{figure}[t]
	\includegraphics[width=3.4in]{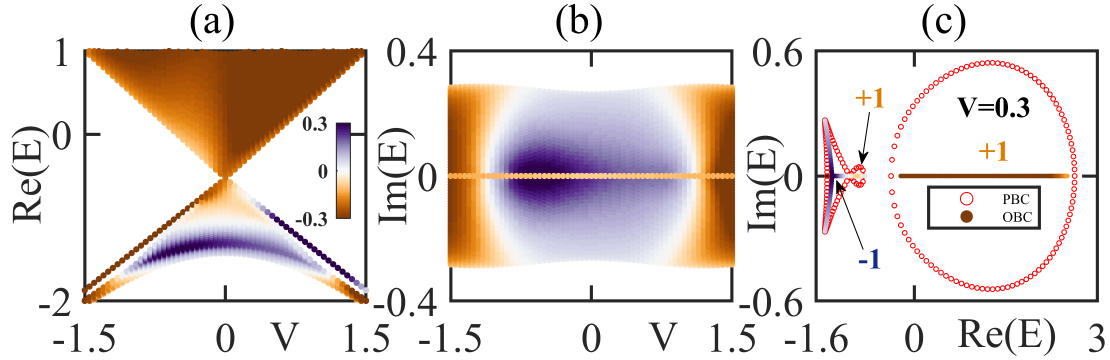}
	\caption{(Color online) Eigenenergy spectrum for the lattices with $\gamma_1=0.1$. Panels (a) and (b) show the real and imaginary parts of the spectrum under OBC as a function of $V$. (c) The spectral structures of under PBC and OBC when $V=0.3$. The colorbar indicates the dIPR values of the eigenstates under OBC and the red dots represent the PBC spectra. Other parameters: $t_1=1$, $t_2=0.3$, $\gamma_2=0.2$, and $N=100$.}
	\label{fig3}
\end{figure}
 
It is well known that the NHSE under OBC is related to the point gap in the PBC spectrum~\cite{Okuma2020PRL,Zhang2020PRL}, which can be characterized by the following winding number
\begin{equation}\label{W}
	W = \frac{1}{2\pi i} \int_{-\pi}^{\pi} dk \frac{d}{dk} \log \det \left[ H(k)-E_B \right],
\end{equation}
where $E_B$ is the base energy. The direction of the NHSE is determined by the sign of the winding number. If $W$ is positive (negative), the corresponding eigenstates under OBC are localized at the left (right) end of the 1D lattice. The directional reversal of the NHSE is related to the sign change of the winding number. So, if the eigenstates of the same band are localized at opposite ends of the lattice, then the PBC spectrum should become twisted and self-crossed. Thus the NHSE edge can be obtained by calculating the self-intersection point in the PBC spectrum. In Fig.~\ref{fig2}, we have marked out the winding number of the loops formed by the PBC spectrum, as indicated by the $\pm 1$ in the figures. It is clear that, for the eigenstates under OBC localized at the opposite ends of the lattice, the sign of the winding number changes from positive to negative or vice versa. Thus we can identify the critical energy that separates the states localized at opposite ends by calculating the self-intersection point in the PBC spectrum. In Figs.~\ref{fig2}(c)-\ref{fig2}(f), we present both the OBC and PBC spectra for different values of $V$. For $V=0$, there is only one band and it is self-crossed as shown in Fig.~\ref{fig2}(c). When $V$ becomes nonzero, the spectrum splits into two separable bands. In Figs.~\ref{fig2}(d) and \ref{fig2}(e), we plot the spectrum of $E_{-}$ and $E_{+}$ for the system with $V=0.2$. There are two self-intersections in the loop enclosing $E_{-}$ and there are NHSE edges. However, for the $E_{+}$ band, there is no self-intersection and the states are all localized at the left end. When $V$ becomes strong enough, there will be no self-intersections for both bands and the NHSE under OBC will be in the left direction, see Fig.~\ref{fig2}(c). For the case with $\gamma_1=0$ discussed here, we can analytically determine the NHSE edge by setting $E_{-}(k_1)=E_{-}(k_2)$ with $k_1\neq k_2$, which leads to
\begin{equation}
	\begin{split}
	&\text{Re}(E_{-}^{SI}) = -\frac{t_1^2}{2 t_2}, \\
	&\text{Im}(E_{-}^{SI}) = \pm 2\gamma_2 \sqrt{1- \frac{4(V^2+2t_1^2)t_2^2-t_1^4}{16t_2^4}}.
	\end{split}
\end{equation}
Notice that here the onsite modulation satisfies the following conditions:
\begin{equation}
	\frac{t_1^4}{4t_2^2} - 2t_1^2 < V^2 < \frac{t_1^4}{4t_2^2} + 4 t_2^2 -2t_1^2.
\end{equation}
Otherwise, there will be no self-intersections in the band $E_{-}(k)$. Thus the NHSE edge in the real part of the spectrum is a constant, while in the imaginary part, it forms an ellipse, as shown by the black lines in Figs.~\ref{fig2}(a) and \ref{fig2}(b). So, by tuning the onsite modulation, we can change the NHSE direction and the position of the NHSE edge.

\begin{widetext}
	
	\begin{figure}[t]
		\includegraphics[width=5.0in]{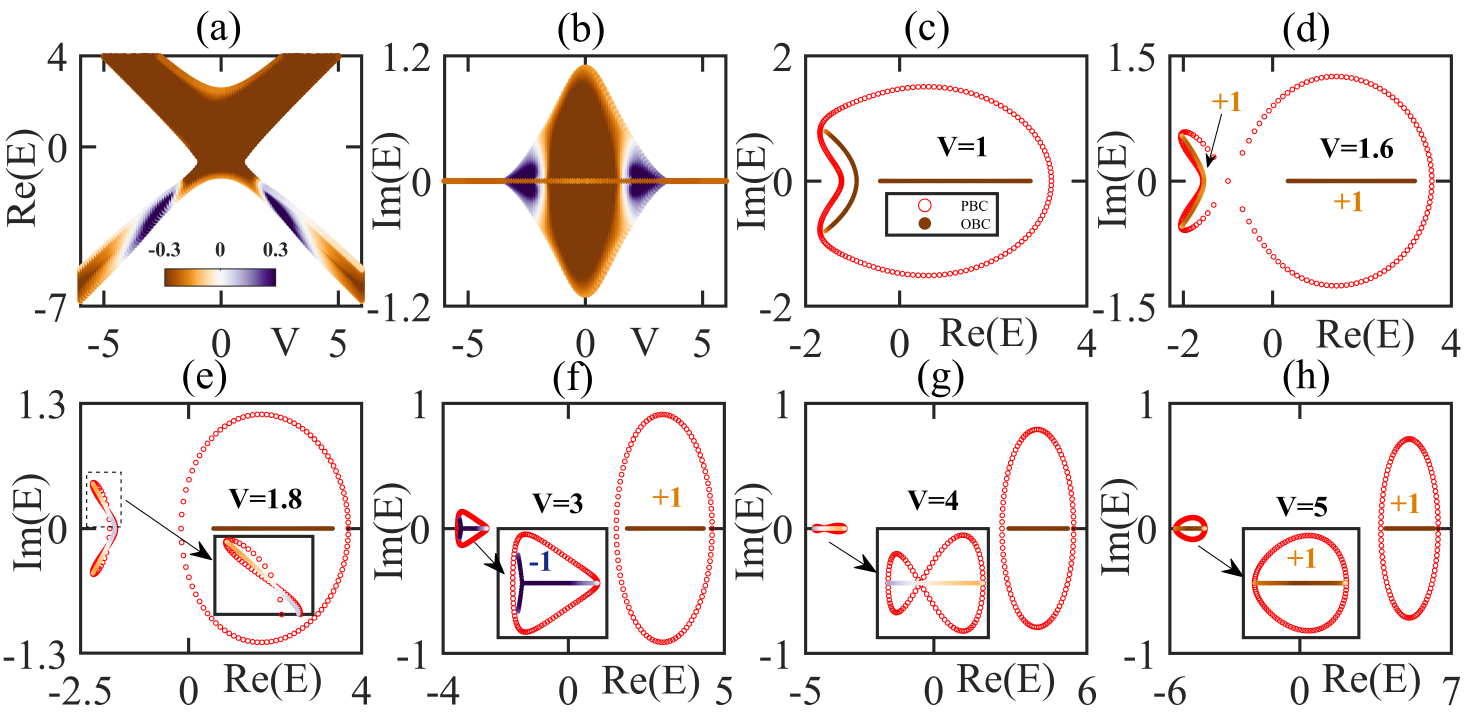}
		\caption{(Color online) Direction reversal of the NHSE under the influence of onsite modulations. Panels (a) and (b) show the real and imaginary parts of the OBC spectrum. The color bar indicates the dIPR values of the eigenstates under OBC. The eigenstates for the lower band $E_{-}$ are localized at the left end of the lattice when $|V|$ is small. As $|V|$ increases, the direction of NHSE is changed to the right direction and back to the left direction in the end. Panels (c)-(h) exhibit the evolution of the spectral structures as the onsite modulation varies from $1$ to $5$. The red dots represent the PBC spectra. The insets are the enlarged view of the band with negative real parts. The numbers $\pm 1$ in the figure are the winding number of the PBC spectrum. Parameters: $t_1=1$, $t_2=0.5$, $\gamma_1=0.8$, $\gamma_2=0.2$, and $N=100$.}
		\label{fig4}
	\end{figure}
	
\end{widetext} 

Next, we turn to the more general cases with $\gamma_1 \neq 0$. Figure \ref{fig3} illustrates the energy spectra for the lattice with $\gamma=0.1$. Since $\gamma_1$ here is relatively small, the behaviors of the NHSE are similar to the above case with $\gamma_1=0$. There are NHSE edges in the lower band $E_{-}$ though they are not constant. The edges will also disappear when $V$ gets strong enough. A more interesting case emerges when $\gamma_1$ gets larger, as shown in Fig.~\ref{fig4}. Here we set $\gamma_1=0.8$. There is no NHSE edge in the spectrum when $V$ is small, as indicated by the dIPR values in Figs.~\ref{fig4}(a) and \ref{fig4}(b). And all the eigenstates are localized at the left end of the 1D lattice. The two energy bands under PBC are inseparable when $|V|<2\gamma_1$ and there are no self-intersections, as shown in Fig.~\ref{fig4}(c). When $|V|=2\gamma_1$, the two bands become degenerated. All the eigenstates still localized at the left end by then; see Fig.~\ref{fig4}(d). As $V$ increases, the NHSE edges will emerge and then disappear in the lower band as the corresponding PBC spectrum $E_{-}(k)$ gets twisted and untwisted, as shown in Figs.~\ref{fig4}(e) and \ref{fig4}(f). Now, all the eigenstates in the $E_{-}$ band are all localized at the right end, while those for the upper band $E_{+}$ are all localized at the left end. Thus the directions of the NHSE for the two bands are opposite. By further increasing the on-site modulation, the lower band $E_{-}(k)$ will become self-crossed again and the NHSE edge reappears in the spectrum; see Fig.~\ref{fig4}(g). Finally, when $V$ becomes strong enough, the loop will be untwisted and the NHSE edge disappears again. Now, the eigenstates in both the bands are localized at the same end, i.e., the left end, as shown in Fig.~\ref{fig4}(h). 

From the above discussions, we can see that the direction of the NHSE can be reversed by tuning the strength of the on-site modulations. The interplay between the on-site modulations and nonreciprocal hopping results in very interesting phenomena in the spectral structures and the behaviors of the NHSE.

\section{Coexistence of extended states and NHSE}\label{Sec4}
We have already shown that due to the interplay between on-site modulation and NNN nonreciprocal hopping, the NHSE can be energy dependent and the direction of the NHSE can be reversed. The eigenstate corresponding to the NHSE edge, i.e., the critical energy that separates the eigenstates localized at the opposite ends of the lattice, is extended, as can be seen from the dIPR values in the above Figs.~\ref{fig4}(a) and \ref{fig4}(b). It will be interesting to check whether more eigenstates, other than the critical states, or even all the eigenstates can be extended in this nonreciprocal lattice. Since the existence of NHSE is closely related to the point gap in the PBC spectrum, if we want to obtain extended eigenstates under OBC, then the point gap in the PBC spectrum needs to be eliminated. At least for those extended eigenstates, the corresponding eigenenergies under PBC should not form a closed loop that encircles the OBC spectrum. From the formula of $E_(k)$ in Eq. ~(\ref{Ek}), we can get a simple case by setting $t_1=t_2=0$. Then the PBC spectrum becomes
\begin{equation}\label{Ek2}
	E_{\pm}(k) = 2 i \gamma_2 \sin k \pm \sqrt{V^2 - 4 \gamma_1^2 \sin^2 \frac{k}{2}}.
\end{equation}
Obviously, if we have no onsite modulations, i.e., $V=0$, the PBC spectrum will become purely imaginary. Then all the eigenstates under OBC are extended. The eigenenergies only move along the imaginary axis and cannot form a loop in the complex energy plane. Thus there is no NHSE in the system. If $0<|V|<2|\gamma_1|$, some of the eigenenergies will become complex with the others still being imaginary. We can expect that there will be some extended states under OBC, while the other states are localized at the ends by NHSE. Thus the extended states could coexist with the eigenstates localized at the ends of the 1D lattice by NHSE. If the on-site modulation further increases to $|V|>2 |\gamma_1|$, then the PBC spectrum would be complex and form loops in the complex energy plane as $k$ varies and there will be no extended states anymore. This is different from the system with only NN hopping [i.e., setting $\gamma_2=0$ in Eq.~(\ref{Ek2})], where all the eigenstates are extended when $|V|<2 |\gamma_1|$ but become localized at the end when $|V|>2 |\gamma_1|$.

\begin{figure}[t]
	\includegraphics[width=3.4in]{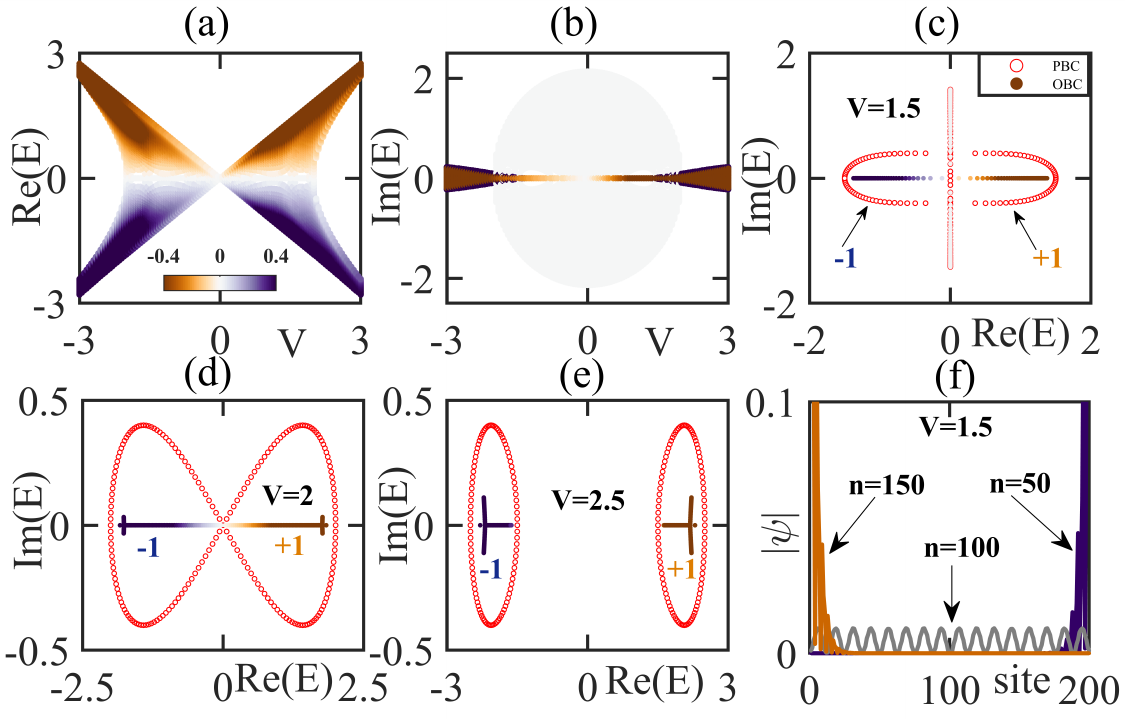}
	\caption{(Color online) Coexistence of extended states and NHSE in the 1D lattice with both on-site modulation and nonreciprocal hopping. Panels (a) and (b) are the real and imaginary parts of the OBC eigenenergies of the system as a function of $V$. Panels (c)-(e) show the PBC and OBC spectra under different on-site modulation. Notice that, in (c), the imaginary eigenenergies under PBC and OBC are overlapped with each other. Panel (f) shows the distributions of the extended states and the states localized at different ends of the lattice. The label $n$ of the eigenstate is obtained by sorting the eigenenergies by their real parts. Parameters: $t_1=0$, $t_2=0$, $\gamma_1=1$, and $\gamma_2=0.2$. The number of unit cells is  $N=100$ and the lattice size is $L=2N=200$.}
	\label{fig5}
\end{figure}

In Fig.~\ref{fig5}, we present the variation of the spectrum as $V$ varies for the case with $t_1=t_2=0$, $\gamma_1=1$, and $\gamma_2=0.2$. The white-shaded region in Figs.~\ref{fig5}(a) and \ref{fig5}(b) indicates that the dIPR values are very close to $0$ for those eigenstates. When $V=1.5$, we can see that some eigenenergies under PBC sit on the imaginary axis while others form loops, as shown in Fig.~\ref{fig5}(c). Notice that the PBC eigenenergies on the imaginary axis are almost identified with their OBC eigenenergies, while the other PBC eigenenergies encircle their corresponding OBC eigenenergies. We have sorted the eigenenergies by the real part and plotted the distribution of the eigenstates under OBC in Fig.~\ref{fig5}(f). Clearly, we have states localized at the left or right end and we also have states extended over the whole lattice. Thus the extended states coexist with NHSE in the same nonreciprocal lattice. Notice that, since $t_1=t_2=0$, the forward NN and NNN hoppings are $-\gamma_1$ and $-\gamma_2$, while the backward NN and NNN hoppings are $\gamma_1$ and $\gamma_2$. So the hoppings along the two directions are of the same strength, only the signs are opposite, and there is no NHSE in the lattice without on-site modulations. It is interesting to show that by introducing on-site modulations, NHSE can be induced and can coexist in the same system with extended states.  By further increasing the value of $V$, such that $0<|V|<2$, the number of extended states will decrease gradually. When $|V|=2$, there is only one extended state, which corresponds to the critical energy that separates the eigenstates localized at the opposite ends; see Fig.~\ref{fig5}(d). Finally, all the extended states will disappear when $|V>2|$, as shown in Fig.~\ref{fig5}(e). Thus the interplay between the nonreciprocal hopping and the real on-site modulations in this case results in the emergence of NHSE and the disappearance of extended states.        

\begin{figure}[t]
	\includegraphics[width=3.4in]{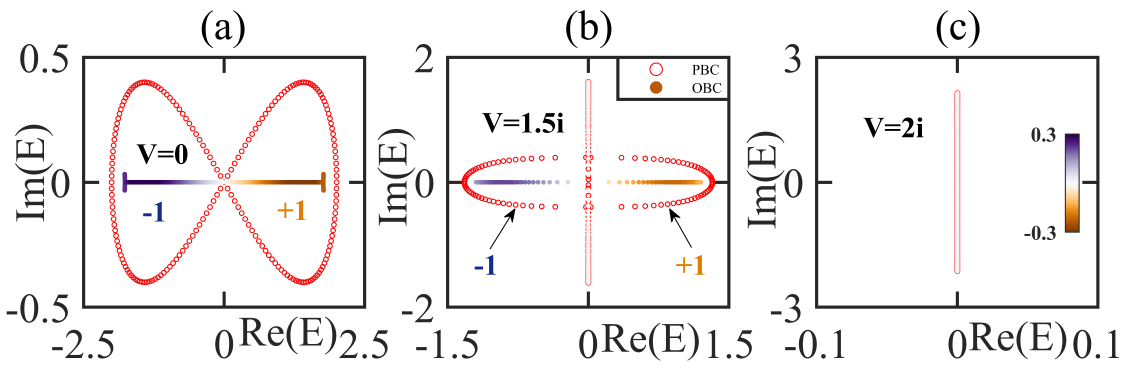}
	\caption{(Color online) Variation of the NHSE under the influences of imaginary onsite modulations: (a) $V=0$, (b) $V=1.5 i$, and (c) $V=2 i$. Parameters: $t_1=1$, $t_2=0$, $\gamma_1=0$, $\gamma_2=0.2$, and $N=100$.}
	\label{fig6}
\end{figure}  

\section{The influences of imaginary modulations}\label{Sec5}
In the previous sections, we have investigated the behaviors of the eigenenergy spectrum and the NHSE under the influences of real on-site modulations. If $V$ becomes imaginary, which represents the physical gain and loss on the lattice site, we can also check the variations in the spectrum and eigenstates using a similar method. The behaviors of the NHSE and the appearance of extended states can also be observed. Most interestingly, we can dissolve the NHSE completely by tuning the strength of the physical gain and loss. In Fig.~\ref{fig6}, we plot the PBC and OBC spectra of the system with different imaginary $V$ values. Other parameters are set as $t_1=1$, $t_2=0$, $\gamma_1=0$, and $\gamma_2=0.2$. Now the variation of the PBC spectrum can be analyzed by the following equation:
\begin{equation}
	E_{\pm}(k) = 2 i \gamma_2 \sin k \pm \sqrt{-V^2 + 4 t_1^2 \cos^2 \frac{k}{2}}.
\end{equation}
When $V=0$, the PBC spectrum can be complex and form a closed loop in the complex energy plane, meaning there will be a point gap in the spectrum. Different from the case in Fig. \ref{fig5}, here the NHSE still shows up when $V=0$, even though the strengths along the forward and backward NNN hopping are the same, as shown in Fig. 6(a). As $V$ increases, the loops formed by the PBC spectrum will shrink. More and more eigenenergies become imaginary, which are overlapped with those under OBC. The eigenstates corresponding to the imaginary energies are extended states, while those with real eigenenergies under OBC localize at the ends of the lattice; see Fig.~\ref{fig6}(b). When $|V| \geq 2$, all the eigenenergies are imaginary and all the states are extended, as shown in Fig.~\ref{fig6}(c). The NHSE thus is completely dissolved by increasing the strength of physical gain and loss. The role played by the imaginary on-site modulations is quite different from the real on-site modulations discussed in Sec.~\ref{Sec4}, where the increasing of $V$ will eliminate the extended states. Here, on the contrary, we find more and more extended eigenstates when the imaginary modulation grows and all the states become extended in the end. These results illustrate the different influences of the real modulations and physical gain/loss in the nonreciprocal systems. The disappearance of NHSE has also been reported in a recent work~\cite{Halder2024PRB}, where a two-orbital non-Hermitian 1D chain with nonreciprocal hopping is considered.

Comparing with the case shown in Fig.~\ref{fig2} with only several extended states showing up, which correspond to the self-intersecting points in the PBC spectrum, here by setting $t_2=\gamma_1=0$ (or by setting $t_1=t_2=0$ as in Sec.~\ref{Sec4}), we can obtain a large number of extended states as we tune the real or imaginary onsite modulations.

\section{Summary}\label{Sec6}
In summary, we have investigated the spectral properties and the behaviors of the NHSE in the 1D lattices with staggered on-site modulations and nonreciprocal hoppings. We find that the interplay between the on-site modulation and the nonreciprocity in the nearest- and next-nearest-neighboring hopping leads to a variety of the exotic properties in such non-Hermitian systems, which cannot be observed in systems without on-site modulations or with only nearest-neighboring nonreciprocal hopping. The variation of the on-site modulation could modify the direction of the NHSE and the position of the NHSE edge in the system. Moreover, we find that extended states can coexist with NHSE in the same system and the NHSE can be dissolved completely if the onsite modulations are imaginary. As to the systems with longer-range nonreciprocal hopping, it has been shown that NHSE edges also exist~\cite{Zeng2022PRB2}. Similar to the model with next-nearest-neighboring hopping discussed in this work, we can also expect the emergence of a large number of extended states which coexist with the NHSE by further introducing on-site modulations. Our work unveils the intricate interplay between on-site modulations and nonreciprocal hopping in non-Hermitian systems.

\begin{acknowledgments}
This work is supported by the National Natural Science Foundation of China (Grant No. 12204326) and Beijing Natural Science Foundation (Grant No. 1232030).
\end{acknowledgments}


\begin{thebibliography}{}
\bibitem{Bender1998PRL}{C. M. Bender and S. Boettcher, \href{https://doi.org/10.1103/PhysRevLett.80.5243}{Phys. Rev. Lett. \textbf{80,} 5243 (1998).}}

\bibitem{Bender2002PRL}{C. M. Bender, D. C. Brody, and H. F. Jones, \href{https://doi.org/10.1103/PhysRevLett.89.270401}{Phys. Rev. Lett. \textbf{89,} 270401 (2002).}}

\bibitem{Bender2007RPP}{C. M. Bender, \href{https://doi.org/10.1088/0034-4885/70/6/R03}{Rep. Prog. Phys. \textbf{70,} 947 (2007).}}

\bibitem{Mostafazadeh2002JMP}{A. Mostafazadeh, \href{https://doi.org/10.1063/1.1418246}{J. Math. Phys. \textbf{43,} 205 (2002);} \href{ https://doi.org/10.1063/1.1461427}{\emph{ibid.} \textbf{43,} 2814 (2002);} \href{https://doi.org/10.1063/1.1489072}{\emph{ibid.} \textbf{43,} 3944 (2002).}}

\bibitem{Mostafazadeh2010IJMMP}{A. Mostafazadeh, \href{https://doi.org/10.1142/S0219887810004816}{Int. J. Geom. Meth. Mod. Phys. \textbf{7,} 1191 (2010).}}

\bibitem{Moiseyev2011Book}{N. Moiseyev, \emph{Non-Hermitian Quantum Mechanics} (Cambridge University Press, Cambridge, UK, 2011).}

\bibitem{Zeng2020PRB1}{Q.-B. Zeng, Y.-B. Yang, and R. L\"u, \href{https://doi.org/10.1103/PhysRevB.101.125418}{Phys. Rev. B \textbf{101,} 125418 (2020).}}

\bibitem{Kawabata2020PRR}{K. Kawabata and M. Sato, \href{https://doi.org/10.1103/PhysRevResearch.2.033391}{Phys. Rev. Research \textbf{2,} 033391 (2020).}}

\bibitem{Zeng2021NJP}{Q.-B. Zeng and Rong L\"u, \href{https://doi.org/10.1088/1367-2630/ac61d0}{New J. Phys. \textbf{24,} 043023 (2022).}}

\bibitem{Cao2015RMP}{H. Cao and J. Wiersig, \href{https://doi.org/10.1103/RevModPhys.87.61}{Rev. Mod. Phys. \textbf{87,} 61 (2015).}}

\bibitem{Konotop2016RMP}{V. V. Konotop, J. Yang, and D. A. Zezyulin, \href{https://doi.org/10.1103/RevModPhys.88.035002}{Rev. Mod. Phys. \textbf{88,} 035002 (2016).}}

\bibitem{Ganainy2018NatPhy}{R. El-Ganainy, K. G. Makris, M. Khajavikhan, Z. H. Musslimani, S. Rotter, and D. N. Christodoulides, \href{https://doi.org/10.1038/nphys4323}{Nat. Phys. \textbf{14,} 11 (2018).}}

\bibitem{Ashida2020AiP}{Y. Ashida, Z. Gong, and M. Ueda, \href{https://doi.org/10.1080/00018732.2021.1876991}{Advances in Physics \textbf{69,} 249 (2020).}}

\bibitem{Bergholtz2021RMP}{E. J. Bergholtz, J. C. Budich, and F. K. Kunst, \href{https://doi.org/10.1103/RevModPhys.93.015005}{Rev. Mod. Phys. \textbf{93,} 015005 (2021).}}	

\bibitem{Makris2008PRL}{K. G. Makris, R. El-Ganainy, D. N. Christodoulides, and Z. H. Musslimani, \href{https://doi.org/10.1103/PhysRevLett.100.103904}{Phys. Rev. Lett. \textbf{100,} 103904 (2008).}}

\bibitem{Klaiman2008PRL}{S. Klaiman, U. Günther, and N. Moiseyev, \href{https://doi.org/10.1103/PhysRevLett.101.080402}{Phys. Rev. Lett. \textbf{101,} 080402 (2008).}}

\bibitem{Guo2009PRL}{A. Guo, G. J. Salamo, D. Duchesne, R. Morandotti, M. VolatierRavat, V. Aimez, G. A. Siviloglou, and D. N. Christodoulides, \href{https://doi.org/10.1103/PhysRevLett.103.093902}{Phys. Rev. Lett. \textbf{103,} 093902 (2009).}}

\bibitem{Ruter2010NatPhys}{C. E. Rüter, K. G. Makris, R. El-Ganainy, D. N. Christodoulides, M. Segev, and D. Kip, \href{https://doi.org/10.1038/nphys1515}{Nat. Phys. \textbf{6,} 192 (2010).}}

\bibitem{Lin2011PRL}{Z. Lin, H. Ramezani, T. Eichelkraut, T. Kottos, H. Cao, and D. N. Christodoulides, \href{https://doi.org/10.1103/PhysRevLett.106.213901}{Phys. Rev. Lett. \textbf{106,} 213901 (2011).}}

\bibitem{Regensburger2012Nat}{A. Regensburger, C. Bersch, M.-A. Miri, G. Onishchukov, D. N. Christodoulides, and U. Peschel, \href{https://doi.org/10.1038/nature11298}{Nature (London) \textbf{488,} 167 (2012).}}

\bibitem{Feng2013NatMat}{L. Feng, Y.-L. Xu, W. S. Fegadolli, M.-H. Lu, J. E. B. Oliveira, V. R. Almeida, Y.-F. Chen, and A. Scherer, \href{https://doi.org/10.1038/nmat3495}{Nat. Mater. \textbf{12,} 108 (2013).}}

\bibitem{Peng2014NatPhys}{B. Peng, S. K. Özdemir, F. Lei, F. Monifi, M. Gianfreda, G. L. Long, S. Fan, F. Nori, C. M. Bender, and L. Yang, \href{https://doi.org/10.1038/nphys2927}{Nat. Phys. \textbf{10,} 394 (2014).}}

\bibitem{Wiersig2014PRL}{J. Wiersig, \href{https://doi.org/10.1103/PhysRevLett.112.203901}{Phys. Rev. Lett. \textbf{112,} 203901 (2014).}}

\bibitem{Hodaei2017Nat}{H. Hodaei, A. U. Hassan, S. Wittek, H. Garcia-Gracia, R. El-Ganainy, D. N. Christodoulides, and M. Khajavikhan, \href{https://doi.org/10.1038/nature23280}{Nature (London) \textbf{548,} 187 (2017).}}

\bibitem{Chen2017Nat}{W. Chen, ¸ S. K. Özdemir, G. Zhao, J. Wiersig, and L. Yang, \href{https://doi.org/10.1038/nature23281}{Nature (London) \textbf{548,} 192 (2017).}}

\bibitem{Brody2012PRL}{D. C. Brody and E.-M. Graefe, \href{https://doi.org/10.1103/PhysRevLett.109.230405}{Phys. Rev. Lett. \textbf{109,} 230405 (2012).}}

\bibitem{Lee2014PRX}{T. E. Lee and C.-K. Chan, \href{https://doi.org/10.1103/PhysRevX.4.041001}{Phys. Rev. X \textbf{4,} 041001 (2014).}}

\bibitem{Li2019NatCom}{J. Li, A. K. Harter, J. Liu, L. de Melo, Y. N. Joglekar, and L. Luo, \href{https://doi.org/10.1038/s41467-019-08596-1}{Nat. Commun. \textbf{10,} 855 (2019).}}

\bibitem{Kawabata2017PRL}{K. Kawabata, Y. Ashida, and M. Ueda, \href{https://doi.org/10.1103/PhysRevLett.119.190401}{Phys. Rev. Lett. \textbf{119,} 190401 (2017).}}

\bibitem{Hamazaki2019PRL}{R. Hamazaki, K. Kawabata, and M. Ueda, \href{https://doi.org/10.1103/PhysRevLett.123.090603}{Phys. Rev. Lett. \textbf{123,} 090603 (2019).}}

\bibitem{Xiao2019PRL}{L. Xiao, K. Wang, X. Zhan, Z. Bian, K. Kawabata, M. Ueda, W. Yi, and P. Xue, \href{https://doi.org/10.1103/PhysRevLett.123.230401}{Phys. Rev. Lett. \textbf{123,} 230401 (2019).}}

\bibitem{Wu2019Science}{Y. Wu, W. Liu, J. Geng, X. Song, X. Ye, C.-K. Duan, X. Rong, and J. Du, \href{https://doi.org/10.1126/science.aaw8205}{Science \textbf{364,} 878 (2019).}}

\bibitem{Yamamoto2019PRL}{K. Yamamoto, M. Nakagawa, K. Adachi, K. Takasan, M. Ueda, and N. Kawakami, \href{https://doi.org/10.1103/PhysRevLett.123.123601}{Phys. Rev. Lett. \textbf{123,} 123601 (2019).}}

\bibitem{Naghiloo2019NatPhys}{M. Naghiloo, N. Abbasi, Y. N. Joglekar, and K. W. Murch, \href{https://doi.org/10.1038/s41567-019-0652-z}{Nat. Phys. \textbf{15,} 1232 (2019).}}

\bibitem{Matsumoto2020PRL}{N. Matsumoto, K. Kawabata, Y. Ashida, S. Furukawa, and M. Ueda, \href{https://doi.org/10.1103/PhysRevLett.125.260601}{Phys. Rev. Lett. \textbf{125,} 260601 (2020).}}

\bibitem{Mandal2020PRL}{S. Mandal, R. Banerjee, Elena A. Ostrovskaya, and T. C. H. Liew, \href{https://doi.org/10.1103/PhysRevLett.125.123902}{Phys. Rev. Lett. \textbf{125,} 123902 (2020).}}

\bibitem{Mandal2022ACS}{S. Mandal, R. Banerjee, and T. C. H. Liew, \href{https://doi.org/10.1021/acsphotonics.1c01425}{ACS Photon. \textbf{9,} 527 (2022).} }

\bibitem{Yao2018PRL1}{S. Yao and Z. Wang, \href{https://doi.org/10.1103/PhysRevLett.121.086803}{Phys. Rev. Lett. \textbf{121,} 086803 (2018).}}

\bibitem{Yao2018PRL2}{S. Yao, F. Song, and Z. Wang, \href{https://doi.org/10.1103/PhysRevLett.121.136802}{Phys. Rev. Lett. \textbf{121,} 136802 (2018).}}

\bibitem{Alvarez2018PRB}{V. M. Martinez Alvarez, J. E. Barrios Vargas, and L. E. F. Foa Torres, \href{https://doi.org/10.1103/PhysRevB.97.121401}{Phys. Rev. B \textbf{97,} 121401(R) (2018).}}

\bibitem{Alvarez2018EPJ}{V. M. Martinez Alvarez, J. E. Barrios Vargas, M. Berdakin, and L. E. F. Foa Torres, \href{https://doi.org/10.1140/epjst/e2018-800091-5}{Eur. Phys. J. Spec. Top. \textbf{227,} 1295 (2018).}}

\bibitem{Lee2019PRB}{C. H. Lee and R. Thomale, \href{https://journals.aps.org/prb/abstract/10.1103/PhysRevB.99.201103}{Phys. Rev. B \textbf{99,} 201103(R) (2019).}}

\bibitem{Zhou2019PRB}{H. Zhou and J. Y. Lee, \href{https://doi.org/10.1103/PhysRevB.99.235112}{Phys. Rev. B \textbf{99,} 235112 (2019).}}

\bibitem{Kawabata2019PRX}{K. Kawabata, K. Shiozaki, M. Ueda, and M. Sato, \href{https://doi.org/10.1103/PhysRevX.9.041015}{Phys. Rev. X \textbf{9,} 041015 (2019).}}

\bibitem{Song2019PRL}{F. Song, S. Yao, and Z. Wang, \href{https://doi.org/10.1103/PhysRevLett.123.170401}{Phys. Rev. Lett. \textbf{123,} 170401 (2019).}}

\bibitem{Okuma2020PRB}{N. Okuma and Masatoshi Sato, \href{https://doi.org/10.1103/PhysRevB.102.014203}{Phys. Rev. B \textbf{102,} 014203 (2020).}}

\bibitem{Xiao2020NatPhys}{L. Xiao, T. Deng, K. Wang, G. Zhu, Z. Wang, W. Yi, and P. Xue, \href{https://doi.org/10.1038/s41567-020-0836-6}{Nat. Phys. \textbf{16,} 761 (2020).}}

\bibitem{Yoshida2020PRR}{T. Yoshida, T. Mizoguchi, and Y. Hatsugai, \href{https://doi.org/10.1103/PhysRevResearch.2.022062}{Phys. Rev. Research \textbf{2,} 022062(R) (2020).}}

\bibitem{Longhi2019PRR}{S. Longhi, \href{https://doi.org/10.1103/PhysRevResearch.1.023013}{Phys. Rev. Research \textbf{1,} 023013 (2019).}}

\bibitem{Yi2020PRL}{Y. Yi and Z. Yang, \href{https://doi.org/10.1103/PhysRevLett.125.186802}{Phys. Rev. Lett. \textbf{125,} 186802 (2020).}}

\bibitem{Claes2021PRB}{J. Claes and T. L. Hughes, \href{https://doi.org/10.1103/PhysRevB.103.L140201}{Phys. Rev. B \textbf{103,} L140201 (2021).}}

\bibitem{Haga2021PRL}{T. Haga, M. Nakagawa, R. Hamazaki, and M. Ueda, \href{https://doi.org/10.1103/PhysRevLett.127.070402}{Phys. Rev. Lett. \textbf{127,} 070402 (2021).}}

\bibitem{Zeng2022PRA}{Q.-B. Zeng and R. L\"u, \href{https://doi.org/10.1103/PhysRevA.105.042211}{Phys. Rev. A \textbf{105,} 042211 (2022).}} 

\bibitem{Zeng2022PRB}{Q.-B. Zeng and R. L\"u, \href{https://doi.org/10.1103/PhysRevB.105.245407}{Phys. Rev. B \textbf{105,} 245407 (2022).}}

\bibitem{Kunst2018PRL}{F. K. Kunst, E. Edvardsson, J. C. Budich, and E. J. Bergholtz, \href{https://doi.org/10.1103/PhysRevLett.121.026808}{Phys. Rev. Lett. \textbf{121,} 026808 (2018).}}

\bibitem{Jin2019PRB}{L. Jin and Z. Song, \href{https://doi.org/10.1103/PhysRevB.99.081103}{Phys. Rev. B \textbf{99,} 081103(R) (2019).}}

\bibitem{Yokomizo2019PRL}{K. Yokomizo and S. Murakami, \href{https://doi.org/10.1103/PhysRevLett.123.066404}{Phys. Rev. Lett. \textbf{123,} 066404 (2019).}}

\bibitem{Herviou2019PRA}{L. Herviou, J. H. Bardarson, and N. Regnault, \href{https://doi.org/10.1103/PhysRevA.99.052118}{Phys. Rev. A \textbf{99,} 052118 (2019).}}

\bibitem{Zeng2020PRB}{Q.-B. Zeng, Y.-B. Yang, and Y. Xu, \href{https://doi.org/10.1103/PhysRevB.101.020201}{Phys. Rev. B 101, 020201(R) (2020).}}

\bibitem{Borgnia2020PRL}{D. S. Borgnia, A. J. Kruchkov, and R.-J. Slager, \href{https://doi.org/10.1103/PhysRevLett.124.056802}{Phys. Rev. Lett. \textbf{124,} 056802 (2020).}}

\bibitem{Yang2020PRL2}{Z. Yang, K. Zhang, C. Fang, and J. Hu, \href{https://doi.org/10.1103/PhysRevLett.125.226402}{Phys. Rev. Lett. \textbf{125,} 226402 (2020).}}

\bibitem{Zirnstein2021PRL}{H.-G. Zirnstein, G. Refael, and B. Rosenow, \href{https://doi.org/10.1103/PhysRevLett.126.216407}{Phys. Rev. Lett. \textbf{126,} 216407 (2021).}}

\bibitem{Zhang2022arxiv}{Z. Q. Zhang, H. Liu, H. Liu, H. Jiang, and X. C. Xie, \href{https://doi.org/10.48550/arXiv.2201.01577}{arXiv:2201.01577.}}

\bibitem{Xiong2018JPC}{Y. Xiong, \href{https://doi.org/10.1088/2399-6528/aab64a}{J. Phys. Commun. \textbf{2,} 035043 (2018).}}

\bibitem{Budich2020PRL}{J. C. Budich and E. J. Bergholtz, \href{https://doi.org/10.1103/PhysRevLett.125.180403}{Phys. Rev. Lett. \textbf{125,} 180403 (2020).}}

\bibitem{Koch2022PRR}{F. Koch and J. C. Budich, \href{https://doi.org/10.1103/PhysRevResearch.4.013113}{Phys. Rev. Research \textbf{4,} 013113 (2022).}}

\bibitem{Okuma2020PRL}{N. Okuma, K. Kawabata, K. Shiozaki, and M. Sato, \href{https://doi.org/10.1103/PhysRevLett.124.086801}{Phys. Rev. Lett. \textbf{124,} 086801 (2020).}}

\bibitem{Zhang2020PRL}{K. Zhang, Z. Yang, and C. Fang, \href{https://doi.org/10.1103/PhysRevLett.125.126402}{Phys. Rev. Lett. \textbf{125,} 126402 (2020).}}

\bibitem{Hatano1996PRL}{N. Hatano and D. R. Nelson, \href{https://doi.org/10.1103/PhysRevLett.77.570}{Phys. Rev. Lett. \textbf{77,} 570 (1996).}}

\bibitem{Shnerb1998PRL}{N. M. Shnerb and D. R. Nelson, \href{https://doi.org/10.1103/PhysRevLett.80.5172}{Phys. Rev. Lett. \textbf{80,} 5172 (1998).}}

\bibitem{Gong2018PRX}{Z. Gong, Y. Ashida, K. Kawabata, K. Takasan, S. Higashikawa, and M. Ueda, \href{https://doi.org/10.1103/PhysRevX.8.031079}{Phys. Rev. X \textbf{8,} 031079 (2018).}}

\bibitem{Jiang2019PRB}{H. Jiang, L.-J. Lang, C. Yang, S.-L. Zhu, and S. Chen, \href{https://doi.org/10.1103/PhysRevB.100.054301}{Phys. Rev. B \textbf{100,} 054301 (2019).}}

\bibitem{Zeng2020PRR}{Q.-B. Zeng and Y. Xu, \href{https://doi.org/10.1103/PhysRevResearch.2.033052}{Phys. Rev. Research \textbf{2,} 033052 (2020).}}

\bibitem{Liu2021PRB1}{Y. Liu, Y. Wang, X. J. Liu, Q. Zhou, and S. Chen, \href{https://doi.org/10.1103/PhysRevB.103.014203}{Phys. Rev. B \textbf{103,} 014203 (2021).}}

\bibitem{Liu2021PRB2}{Y. Liu, Q. Zhou, and S. Chen, \href{https://doi.org/10.1103/PhysRevB.104.024201}{Phys. Rev. B \textbf{104,} 024201 (2021).}}

\bibitem{Zeng2024PRB}{B. Hou, H. Xiao, R. L\"u, and Q.-B. Zeng, \href{https://doi.org/10.1103/PhysRevB.109.094208}{Phys. Rev. B \textbf{109,} 094208 (2024).}}

\bibitem{Zeng2022PRB2}{Q.-B. Zeng, \href{https://doi.org/10.1103/PhysRevB.106.235411}{Phys. Rev. B \textbf{106,} 235411 (2022).}}

%
\bibitem{Halder2024PRB}{D. Halder, R. Thomale, and S. Basu, \href{https://doi.org/10.1103/PhysRevB.109.115407}{Phys. Rev. B \textbf{109,} 115407 (2024).}}

\end{thebibliography}
\end{document}